# Geospatial sustainability assessment of universal Fiber-To-The-Neighborhood (FTTnb) broadband infrastructure strategies for Sub-Saharan Africa

OGUTU B. OSORO[1], EDWARD J OUGHTON[1] and FABION KAUKER[2]

[1] Department of Geography & Geoinformation Science Department, George Mason University, Fairfax, VA 22030, USA.

[2] Pozibl Inc, San Jose, CA 95125, USA.

## Abstract

Broadband Internet access is an important way to help achieve the Sustainable Development Goals. Currently, fixed fiber infrastructure is essential for providing universal broadband, but has received relatively little research attention in low-income countries compared to other more cost-efficient wireless technologies. Yet, pushing out fiber broadband network to local areas is essential, even if the final access network is still wireless. Here, we design least-cost Fiber-To-The-Neighborhood (FTTnb) architectures using two spatial optimization Steiner Tree algorithms to jointly determine investment costs, environmental emissions, and Social Carbon Costs. We find that the average annualized per user emissions in low population density areas (<9 people per $km^2$) range from 0.18-9.6 kg $CO_2$ eq./user, compared to 0.015-0.12 kg $CO_2$ eq./user for high population density areas (>958 people per $km^2$). Moreover, Annualized Total Cost of Ownership per user is 12-90 times lower in high population density areas (>958 people per $km^2$) compared to sparsely populated regions (<9 people per $km^2$). Thus, 48% (about 550 million) of the total Sub-Saharan African population live in areas where FTTnb is viable within the next ten years.

## I. Introduction

According to the 2023 International Telecommunication Union (ITU) report, 2.6 billion people are still not connected to the Internet [1]. Even though the new figure is a reduction from the previous (2.7 billion), the quantity of the unconnected population in Africa still stands at around 864 million. Broadband access has a potential of addressing a broad range of Sustainable Development Goals (SDGs) including poverty elimination (SDG1), industry, innovation, and infrastructure (SDG9) and reduced inequalities (SDG10). However, there are several barriers to adoption including a lack of infrastructure, affordability, perceived relevance and digital readiness each contributing to the digital divide [2]. Hitherto, there has been incremental investment in Internet infrastructure in Africa since the landing of the West Africa Submarine Cable link in 2001, which has now resulted in multiple connections across African coastal cities such as Abidjan, Cairo, Lagos, Luanda, Cape Town, Dar es Salaam and Mombasa among others [3] [4].

Due to cost challenges of common Internet connectivity options, fiber penetration is still low, with many users relying on wireless cellular services. Indeed, Mobile Network Operators (MNOs) have been at the forefront of providing Internet services through mobile phones. For instance, Glo, an MNO in West Africa charges US$ 1.28 for 3.9 GB in Nigeria [5], MTN charges US$ 2.38 for 5 GB in Zambia [6], Airtel charges US$ 0.8 for 2.5 GB in Democratic Republic of Congo [7] and Safaricom in Kenya charges US$ 6.78 for 5 GB [8]. Mobile wireless technologies, such as 4G, are generally cost effective for Sub-Saharan Africa (SSA), with anticipated annual savings of between 7-15% for 10 GB/month compared to previous technologies [9]. However, these services are only available in areas where mobile coverage is present. Since MNOs are rational profit-driven businesses, they have been targeting densely populated urban areas where the Average Revenue per User (ARPU) is high and can enable them to reduce operational costs [10]. Globally, MNOs have been experiencing a challenging business environment, with declining or static ARPU. For instance, the MNO ARPU has dropped by 0.12% from 2022 to 2023 [11], while inflation has been elevated to 10% in some economies [12]. Consequently, MNOs tend to prefer infrastructure deployment in densely populated, profitable regions. Unfortunately, about 58% of the African population lives in sparsely populated rural areas [13] where there is a weak business case for deploying MNO communication assets. Part of the future story for connecting Africa involves efficient fiber deployment.

Despite the landing of long-haul optical fiber in specific coastal and interior cities, further deployment has been slow in many instances. Indeed, providing broadband Internet through fiber can be one of the most expensive options, especially for low population density rural and remote areas. Yet fiber is a superior approach in terms of capacity, latency etc., providing the most reliable and future-proof option [14], [15]. As a result of this expense, other novel methods, such as broadband over power lines, has been suggested in rural and remote areas, if existing power infrastructure is already in place [16]. In situations where deploying broadband fiber is expensive, microwave frequencies are often used for backhaul to connect hard-to-reach cell towers to a fiber point of presence [17], [18], [19], with the GSMA estimating more than 80% of sites in SSA are utilizing this approach [20]. Often in locations where building a single infrastructure might be economically unviable, infrastructure sharing strategies have been suggested as ways to significantly reduce the costs of providing broadband [21], [22]. Importantly, over the long-term fiber is the only future-proof approach.

Apart from affordability issues, the environmental sustainability aspects of these connectivity options are also attracting the attention of governments and international institutions. The concern stems from the fact that the infrastructure operated in supporting Internet connectivity requires electricity that produces emissions such as carbon, nitrogen and sulfur oxides [23], [24], [25]. Additionally, deploying any broadband technologies, whether fiber or microwave radio links, requires construction and erection of new infrastructure that need a significant amount of cement and steel, all with an associated carbon footprint. For instance, about 5-10% of the world's anthropogenic carbon (IV) oxide emissions are from cement production [26]. Similarly, steel production accounts for about 6.7% of the global carbon (IV) oxide emissions [27]. Therefore, efforts to lower emissions even in local broadband construction projects can help to reduce environmental impacts from human development.

As a result, the most environmentally friendly ways to connect the unconnected must be identified. Currently, the Information Communication and Technology (ICT) industry accounts for about 3.6% of global Greenhouse Gas (GHG) emissions [28]. Even though long-haul fiber has been deployed in most parts of SSA, pushing out fiber to access networks has been limited, and generally available only in the most densely populated areas [29]. There are emerging efforts but the majority focus on purely technical aspects, such as capacity [30], [31]. Thus, the development of fiber broadband strategies in SSA requires strong geospatial data integration for estimating user densities in targeted uncovered areas, infrastructure asset expansion plans, and the associated cost and environmental impacts. A series of spatial optimization techniques can be used to optimally connect the unconnected rural population while reducing capital expenditure, operational costs and carbon footprints [32]. Against this background, we aim to answer three important research questions in this analysis, articulated as follows:

1) What is the best way to model fixed broadband infrastructure networks using spatial optimization algorithms to reduce investment costs and emissions?
2) How much investment is required in fixed broadband infrastructure to achieve affordable universal connectivity, for example, focusing on a Fiber-to-the-Neighborhood (FTTnb) approach?
3) What are the environmental sustainability implications of broadband infrastructure strategies?

The paper is structured as follows. Next a literature review on the viability of fiber broadband for rural areas is explored followed by a description of the methodology in section III. The results are reported in section IV before discussions made in section V. Finally, the conclusions, study limitations and further research are identified in section VI.

## II.  Literature Review

Broadband Internet infrastructure encompasses several high-speed transmission technologies including fiber optic, coaxial cable, Digital Subscriber Line (DSL), wireless systems and satellite among others. Investing in these fixed broadband technologies may be expensive compared to other wireless options, especially for developing nations. However, there has been a consensus that a well-integrated broadband system can have a positive impact on society and help unlock economic development [15], [33], [34], [35]. As a result, ensuring that everyone is connected to high-capacity broadband Internet is increasingly becoming a priority of governments in both developed and developing economies. Similarly international development institutions such as the World Bank and United Nations have been at the forefront of facilitating broadband access at educational facilities [36]. However, the disparity in the number of connected people between rural and urban areas persists, and is likely to continue over the next decade while new capital is raised and infrastructure incrementally deployed.

### a. Connecting the Unconnected

Several proposals and solutions have been suggested to connect the rural population but geographic remoteness and low population density is a major challenge [37]. These two issues are further compounded by difficulty in providing coverage in areas of market failure while equally reducing carbon footprint e.g., the quantity of carbon (IV) oxide emitted due to an activity by an entity during a process [38], [39], [40], [41], [42]. The problem is worse for network operators in low-income regions, resulting in poor connectivity, particularly in rural areas [43]. Wireless broadband technologies have been proven effective in some rural contexts but annual spectrum fees and high equipment costs can limit their effectiveness in narrowing the digital divide [44]. In other cases, a more operator-neutral solution has been effective in providing coverage, ranging from a wholesale operator [45] through to an infrastructure sharing approach [46]. Most existing studies have focused on deploying wireless broadband access technologies, with little consideration hitherto of the implications of fixed fiber infrastructure deployment, providing motivation for this paper.

Over the past three decades, fiber optic networks have grown to become the main technology for carrying high quantities of traffic globally and delivering local broadband services to users. However, operators are more likely to deploy fiber networks in urban and suburban locations, compared to rural areas, due to lower deployment costs and higher user densities (leading to greater profitability) [47]. Consequently, deploying rural fiber broadband may require public investment, tax incentives and infrastructure sharing among other strategies to close the digital divide [48]. Even in a frontier broadband market such as South Korea, one evaluation established that $0.81 billion is needed to provide a 100 Mbps Fiber-to-the-Home for all households [49]. Certainly, reusing existing infrastructure in many areas may lower costs. However, this may not be applicable in low-income countries where little existing infrastructure is present, substantially raising the costs of network build-out beyond affordability levels.

Alternative modifications such as routing fiber broadband closest to the consumer (Fiber-to-the-Distribution-Point) have been explored but the results conclude that it is only feasible in those areas with more densely populated premises [50]. In a related study, a delay tolerant network has been proposed that involves transmission of data between separate nodes with microwave backhaul to reduce cost [51]. This opens a research gap for developing optimization models to initially only select potential distribution points in more economically viable areas. The success of such models will be based on the concept of aggregating users from a similar neighborhood to be served by a central single distribution point before providing the last mile connectivity through cellular or Wi-Fi access [52].

### b.  Environmental Emissions Impacts of Broadband

Due to the growing interest in decarbonizing the telecommunication industry, the deployment of broadband infrastructure extends beyond costs to the carbon footprints from operating such systems. For instance, one assessment puts China's 5G base station emissions at $17\pm5$ million metric tonnes of carbon alone in 2020 [53]. As the world migrates to higher capacity wireless broadband, more emissions are expected due to network densification. Higher capacity systems require construction of more radio sites and extra power for operation [54]. Some studies have focused on carbon footprints of dense networks such as 4G/5G [55], including proposing the usage of solar power systems to lower emissions [25]. Preliminary work has shown that fiber broadband has lower associated operational emissions when compared to wireless systems [56]. However, limited studies exist on quantifying the carbon footprint of fiber broadband via different network design options to limit emissions.

Fixed broadband networks have important environmental sustainability dimensions, and are a key part of the overall environmental impact of the ICT sector. However, the discussion on the contribution of ICT technologies (such as fixed networks) to GHG emissions has drawn inconclusive results. For instance, some studies support that ICT indirectly mitigates GHG emissions [57], [58], [59], [60], [61]. While, other contrasting studies indicate that ICT increases electricity consumption, pushing up global GHG emissions overall [40], [62], [63], [64]. For example, the growth of data centers has led to these assets accounting for 1% of global electricity consumption, driven by increased utilization of high definition video, and increasingly the training of artificial intelligence models [65]. In the context of broadband provision, the quality and method (fiber, fixed wireless or satellite) of providing Internet services dictates the associated amount of GHG emissions.

### c.  Network Design Approaches and Life Cycle Assessment (LCA)

Infrastructure networks are inherently spatial, and thus spatial optimization models can be applied in maximizing the number of connected people in rural areas at a lower cost and with fewer emissions. These types of algorithms are used in dynamic applications, such as transportation systems, to establish the most efficient routes at lowest cost [66]. In satellite communication, Monte Carlo simulation techniques combined with spatial optimization algorithms have been used in minimizing the number of ground stations that provide maximum coverage, data rates and availability [67]. A similar approach can be applied in the context of fiber broadband to establish the most optimum location of a central node for serving a geographical area with given population density threshold, at a lower cost and with fewer emissions [47].

Since it has been established that the cost of deploying fiber in rural areas is challenging [44], [68], the focus of spatial optimization algorithms should be on lowering costs and emissions, while maximizing coverage [69]. Usually, high costs and emissions arise from sparse population distributions in potential user areas but studies have shown that mathematical models can be implemented to estimate designs which lower costs [70]. However, the majority of such models focus on technical aspects such as bandwidth and capacity (neglecting emissions) [71].

A key challenge is accounting for the potential locations of network assets given population density distributions, while considering costs and environmental impacts. One option is utilizing Steiner Trees (STs). Some examples of STs are Minimum Spanning Tree (MST) and Prize Collecting Steiner Tree (PCST) algorithms. MST and PCST are a set of spatial optimization algorithms that can be used to design fiber networks, to reduce costs, emissions and potentially maximizing coverage [72]. When combined with clustering algorithms to group areas of demand [73], STs can be useful in identifying the shortest distance between densely populated areas to reduce the quantity of materials needed (in the case of broadband deployment). Although the ST-based models have uncertainties [74], they are useful in designing optimal networks for essential services.

Fiber broadband is one of the most energy efficient technologies with low associated GHG emissions due to two main reasons [75]. Firstly, the silicon (IV) oxide material used to make glass is hugely abundant, requiring less extraction. Secondly, using low-power electronic equipment, fiber has very low attenuated signal loss compared to other technologies. This also results in fewer amplifiers and intermediate devices, and lower operational emissions. Consequently, the use of fiber can significantly lower incurred GHGs. For instance, the European Commission report indicated that 50 Mbps fiber connections emitted 1.7 tons of carbon (IV) oxide in a year compared to 2.7 tons for coaxial cable (due to higher attenuated signal losses over copper) [75].

Life Cycle Assessment (LCA) is a technique that can be applied to model the environmental impact of a product or service over a particular lifetime [76], such as quantifying emissions generated via heterogenous Internet infrastructure architectures. The LCA process should account for emissions due to raw material acquisition, manufacturing, deployment of the network, use of the network, maintenance, repairs and End-of-Life Treatment (EOLT) for the devices [77]. Moreover, LCA should also account for emissions from across the network hierarchy. Fiber broadband provision, like other Internet services, can be split into three domains including the access, edge and core network components [78]. All three domains require accurately profiling to obtain emissions per subscriber.

While fiber broadband has been established as one of the greenest forms of Internet connectivity, some studies have shown that production of a kilometer optical fiber cable can still lead to 6.5 kg of carbon (IV) oxide emissions [28]. Therefore, the length of optical fiber used in connecting users is vital in estimating the associated GHG emissions. The quantification of this material can be related to emissions factors to establish the emissions per connection. Environmental institutions, such as the European Environment Agency, provide emission factors for every material and processes annually in units, such as kilograms of carbon (IV) oxide equivalents (kg $CO_2$ eq.) [79]. Attributing these emission factors to every domain of a fiber network can be challenging but is possible via scenario analysis.

The calculation of total emissions for an infrastructure network, such as fiber broadband, requires aggregation of all emission sources at different stages of the LCA [80]. Identifying and quantifying the type and bill of materials that goes into each of the phases can be challenging. However, data from environmental impact assessments, field surveys, related databases, equipment manuals, official statistics and literature can be used in the analytical process [81], [82]. Although there are uncertainties present within any LCA model, these challenges can be overcome through modeling and simulation techniques, such as Monte Carlo and sensitivity analysis. Now that a thorough review of the literature has been carried out, a method capable of answering the research questions can be specified.

## III.   Methodology

A general theoretical model for providing FTTnb is now defined, consisting of the demand, supply, cost, spatial optimization and emissions components. In this model, we quantify the environmental carbon emissions and costs incurred in building a FTTnb network to local areas. Once the terminal access node has been built, users (households) can connect via other wireless technologies (e.g., 4G, 5G or Wi-Fi). The emissions and costs are not quantified for Fiber-to-the-Premise (FTTP), as we know *a priori* that this architecture is prohibitively expensive for low-income countries. Later this theoretical model is populated with empirical data. The proposed FTTnb network outlay based on the standard ITU guide [83] is shown in Fig. 1.

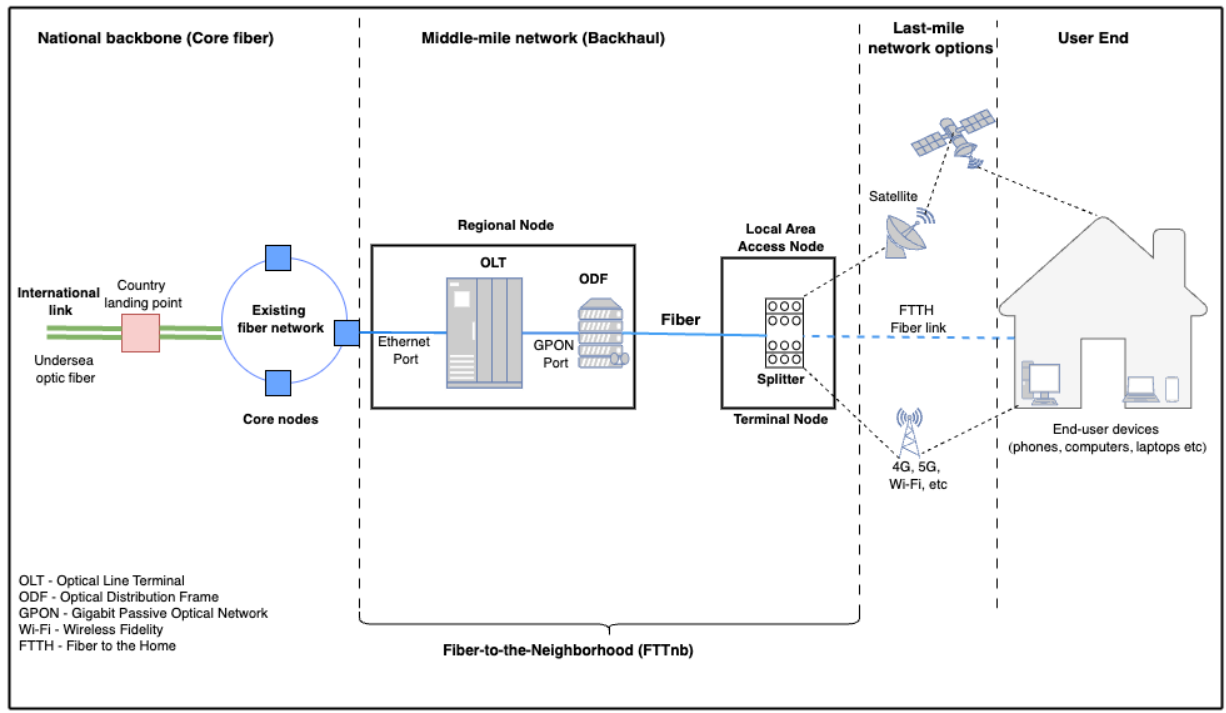

**Fig. 1 | Network architecture.** Fiber-to-the-Neighborhood (FTTnb).

### a.   Demand Model

This model estimates the number and location of the potential users to be connected in an area. First, only areas with population above a set population density threshold ($Pop_{min}$) are identified since it will be necessary to connect remote areas via satellite (due to poor economic viability of terrestrial options). The $Pop_{min}$ can be used to guide the building of nodal networks to the main settlement areas. Considering an area ($A_{(km^2)}$) within a sub-region to be connected by fiber (FiberArea$_i$) with a given population (pop), the population density ($Pop_{(km^2)} > Pop_{min}$) is calculated using equation (1).

$$Pop_{(km^2)} = \frac{pop}{A_{(km^2)}} \qquad (1)$$

The $Pop_{(km^2)}$ is then grouped into ten decile geotypes, where group areas with similar deployment characteristics (e.g., costs) are labeled as Deciles 1 to 10. However, not everyone within the defined $FiberArea_i$ will be connected to the fiber network. Therefore, the model is set to different adoption rates $(AD_r)$ based on a take-up scenario that can be defined as a percentage. For instance, an adoption rate of 0.5% indicates that in a geographic boundary of 10,000 people, only 50 people are likely to be connected to the fiber network. Such an approach is a common way of assessing the demand for infrastructure [84]. Therefore, the number of potential users per square kilometer $(Users_{km^2})$ can then be obtained using equation (2).

$$Users_{km^2} = Pop_{(km^2)} \cdot AD_r \qquad (2)$$

### b. Supply and Spatial Optimization Model

The supply model is made up of the existing fiber nodes from which the deployment begins to future fiber access points. The objective is to connect each of the main population settlement points to the fiber network. A main settlement threshold is defined to help identify where these nodes should be deployed. Since all the settlement population has been estimated, as stated in the demand section, the next step is finding the information on the existing core fiber network $(C_{nod})$. A buffer is created around the existing core fiber network to connect proximate settlements, as operators will choose to connect users close to the existing network before expanding outwards. Secondly, the largest regional settlement with a population above the main settlement threshold is considered as the key routing node for regional fiber access points $(R_{nod})$. Thirdly, the largest sub-regional settlement point is considered as the access nodes $(A_{nod})$. Next, each $A_{nod}$ is connected to each other in a sub-region using Prim's MST and a PCST algorithm to form edges. Each $R_{nod}$ is also connected to each other and back to $C_{nod}$ by the algorithms to obtain the least-cost design (FTTnb). The distance $(d_{(km)})$ calculated based on the two algorithms is used in calculation of costs and emissions.

### c. Cost Model

Constructing FTTnb has an associated cost and carbon emission quantity $(FibTot_{ghg})$ that is a function of the distance $d_{(km)}$. Each algorithm thus minimizes $d_{(km)}$ to reduce costs and $FibTot_{ghg}$ while connecting maximum nodes. The Total Cost of Ownership (TCO) of building the FTTnb is therefore a summation of all the capital expenditure (Capex) and operating expenditure (Opex) over the expected lifetime of the infrastructure (n) at a discount rate, r as shown in equation (3).

$$TCO = Capex + \sum_{y=0}^{n} \frac{Opex}{[1+r]^n} \qquad (3)$$

The Capex consists of Optical Line Terminal (OLT) unit ($C_{OLT}$), civil materials ($C_{Civil}$), transportation ($C_{Trans}$), installation ($C_{Inst}$), Remote Power Unit (RPU) ($C_{RPU}$), Optical Distribution Frame (ODF) ($C_{ODF}$) and splitter ($C_{splt}$), costs summed in equation (4).

$$Capex = C_{OLT} + C_{Civil} + C_{Trans} + C_{Inst} + C_{RPU} + C_{ODF} + C_{splt} \qquad (4)$$

The corresponding Opex costs are site rental ($O_{Rent}$), staff and maintenance ($O_{Staff}$), power ($O_{Pwr}$), regulatory ($O_{Reg}$), customer acquisition ($O_{Acq.}$) and any other ($O_{other}$) costs, equation (5).

$$Opex = O_{Rent} + O_{Staff} + O_{Pwr} + O_{Reg} + O_{Acq.} + O_{other} \qquad (5)$$

### d. Environmental Emissions Model

To quantify the carbon emissions from the FTTnb infrastructure, we follow the LCA framework as defined by the European Telecommunications Standards Institute [77].

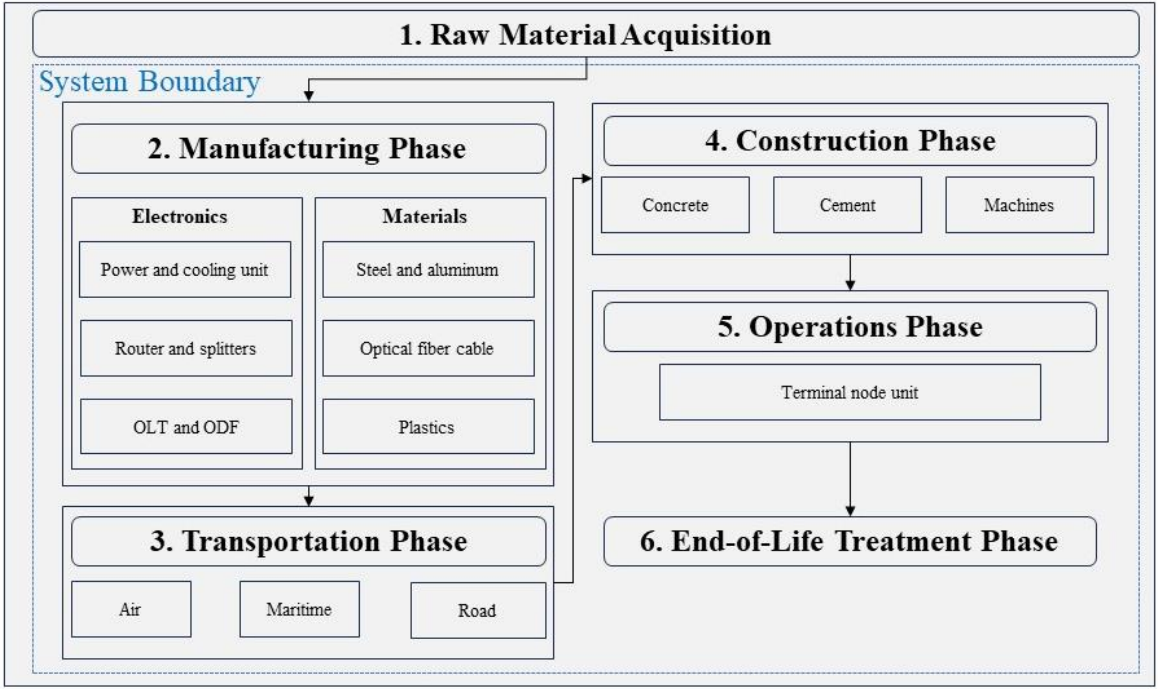

**Fig. 2 | System boundary.** Defined system boundary for LCA of fiber broadband.

The framework outlines raw material acquisition, manufacturing, transportation, construction, operation, and EOLT as the key stages to assessment. In this paper, we consider that the raw materials needed for the network deployment have already been extracted. Hence, the raw material acquisition process is implicitly included in the emission factors of the materials such as steel, aluminum, optical fiber cables, etc. Fig. 2 illustrates the system boundary adopted in this study. Next, the manufacturing phase consists of materials and equipment used in laying out the fiber line links as well as the terminal node. The distance ($d_{(km)}$) between the nodes affects the total amount of fiber optic cable. Given a carbon emission factor of glass, $cf_{gls}$, the total GHG emissions for laying the fiber cable, $fib_{ghg}$ can be calculated using equation (6).

$$\text{fib}_{ghg} = d_{(km)} \times W_{unit} \times cf_{gls} \qquad (6)$$

Where $W_{unit}$, is the weight in kilograms (kg) and $\text{fib}_{ghg}$ is given in kg $CO_2$ eq. The emissions from other non-fiber materials ($\text{nonfib}_{ghg}$) are accounted for using equation (7).

$$\text{nonfib}_{ghg} = \sum_{i=1}^{n} W_{unit} \times cf_i \qquad (7)$$

Where i is the type of the material and $cf_i$, the corresponding carbon emission factor. The total GHG emissions from manufacturing phase, $\text{mfg}_{ghg}$ is thus given by equation (8).

$$\text{mfg}_{ghg} = \text{fib}_{ghg} + \text{nonfib}_{ghg} \qquad (8)$$

For transportation, the resulting emissions are calculated based on the emission factor of material consumption, traveling distance and the mode of transportation. For the OLT, ODF and splitter, the emission values from international shipping travel, $\text{int}_{ghg}$ between overseas plants and the final location port is utilized. For concrete and other materials, the distance between the fiber nodes is used. Therefore, the GHG emissions for transportation of non-fiber materials are calculated using equations (9).

$$\text{nonfb}_{trans} = \sum_{i=1}^{n} W_{unit} \times d_{(km)} \times cf_v \qquad (9)$$

The $cf_v$ is the emission factor of the vehicle fuel used to transport the materials. The total emissions from the transportation phase are thus given by equation (10).

$$\text{trans}_{ghg} = \text{int}_{ghg} + \text{nonfb}_{trans} \qquad (10)$$

In the construction phase, the major emissions are due to trenching and laying of the fiber optic cable. Therefore, the emissions are directly related to the distance covered by the fiber optic cable. However, broadband planning studies have shown that most operators prefer installing the fiber line over the power lines [85]. As a result, only a percentage ($\text{trench}_{\%}$) of the total fiber optic cable is modeled to be placed underground in this study. The total distance trenched ($d_{trench(km)}$) is thus given by equation (11).

$$d_{trench(km)} = d_{(km)} \times \text{trench}_{\%} \qquad (11)$$

Given the quantity of hours needed to operate machinery for every kilometer ($\text{ops}_{hr/km}$), the total operational hours ($\text{ops}_{Total(hrs)}$) and the subsequent total fuel consumption ($\text{Fuel}_{liters}$) can be calculated as in equations (12) and (13), given fuel efficiency ($f_\eta$) value.

$$\text{ops}_{\text{Total(hrs)}} = \text{ops}_{\text{hr/km}} \times d_{\text{trench(km)}} \tag{12}$$

$$\text{Fuel}_{\text{liters}} = \text{ops}_{\text{Total(hrs)}} \times f_{\eta} \tag{13}$$

Given the carbon emission factor of the fuel ($cf_{\text{fuel}}$) used, the resulting construction phase ($\text{Constr}_{\text{ghg}}$) emissions are thus calculated using equation (14).

$$\text{Constr}_{\text{ghg}} = cf_{\text{fuel}} \times \text{Fuel}_{\text{liters}} \tag{14}$$

Next, the GHG emissions from the operation phase are quantified. The electricity power consumption from the central office and terminal node is the main source of emissions at this phase. The power per node metric, $\text{Pn}_{\text{kWh}}$ is used to calculate the final operation emissions. The $\text{Pn}_{\text{kWh}}$ is calculated using the general formula in equation (15) as previously established in the literature [86].

$$\text{Pn}_{\text{kWh}} = \text{P}_{\text{node}} + \left(\frac{\text{P}_{\text{RN}}}{\text{N}_{\text{RN}}}\right) + \alpha\left[\frac{\text{P}_{\text{TU}}}{\text{N}_{\text{TU}}}\right] \tag{15}$$

Where $\text{P}_{\text{node}}$, is the power consumed (all in kilowatts per hour) $\text{P}_{\text{RN}}$, by the fiber node station and $\text{P}_{\text{TU}}$, the terminal node. $\text{N}_{\text{RN}}$ is the number of users sharing the fiber node station and $\text{N}_{\text{TU}}$ the terminal node. The power consumed by the users is not included in the calculation as the system boundary stops at the terminal distribution point. The $\alpha$ value represent additional overheads including electricity distribution losses, external power supplies and cooling needs of the buildings containing the terminal node equipment. The total GHG emissions from the operation phase $\text{ops}_{\text{ghg}}$, is thus estimated by multiplying the total power consumption by the carbon emission factor for electricity usage $cf_{\text{kWh}}$, unit as shown in equation (16).

$$\text{ops}_{\text{ghg}} = \text{Pn}_{\text{kWh}} \times cf_{\text{kWh}} \tag{16}$$

Lastly, the emissions due to EOLT is calculated based on the emission factors from the method of recycling (re-using, open or closed loop cycle recycling). The emission from EOLT of fiber optical cable $\text{fib}_{\text{EOLT}}$, is given by equation (17) and for other non-fiber $\text{nonfib}_{\text{EOLT}}$, materials in equation (18).

$$\text{fib}_{\text{EOLT}} = \text{OLR}_{\text{ghg}} \tag{17}$$

$$\text{nonfib}_{\text{EOLT}} = \sum_{i=1}^{n} \text{Net}_{\text{kg}} \times cf_{i_{\text{rcy}}} \tag{18}$$

Where $\text{OLR}_{\text{ghg}}$, is the emission factor of recycling fiber cable from open loop cycle recycling, $\text{Net}_{\text{kg}}$ the total weight of other materials and $cf_{i_{\text{rcy}}}$, the carbon emission factor for recycling the associated material. The total EOLT emissions is thus given by equation (19).

$$\text{EOLT}_{\text{ghg}} = \text{fib}_{\text{EOLT}} + \text{nonfib}_{\text{EOLT}} \tag{19}$$

The total carbon emissions for the entire fiber LCA are thus given by equation (20).

$$\text{FibTot}_{ghg} = \text{mfg}_{ghg} + \text{trans}_{ghg} + \text{Constr}_{ghg} + \text{ops}_{ghg} + \text{EOLT}_{ghg} \qquad (20)$$

Lastly, the Social Carbon Cost (SCC) is calculated using equation (21) given that 1 tonne of carbon is associated with a cost ($C_{US\$}$) at a discount rate according to the latest SCC research [87].

$$\text{SCC} = \text{FibTot}_{ghg} \times C_{US\$} \qquad (21)$$

An overview of the model is visualized in Fig. 3.

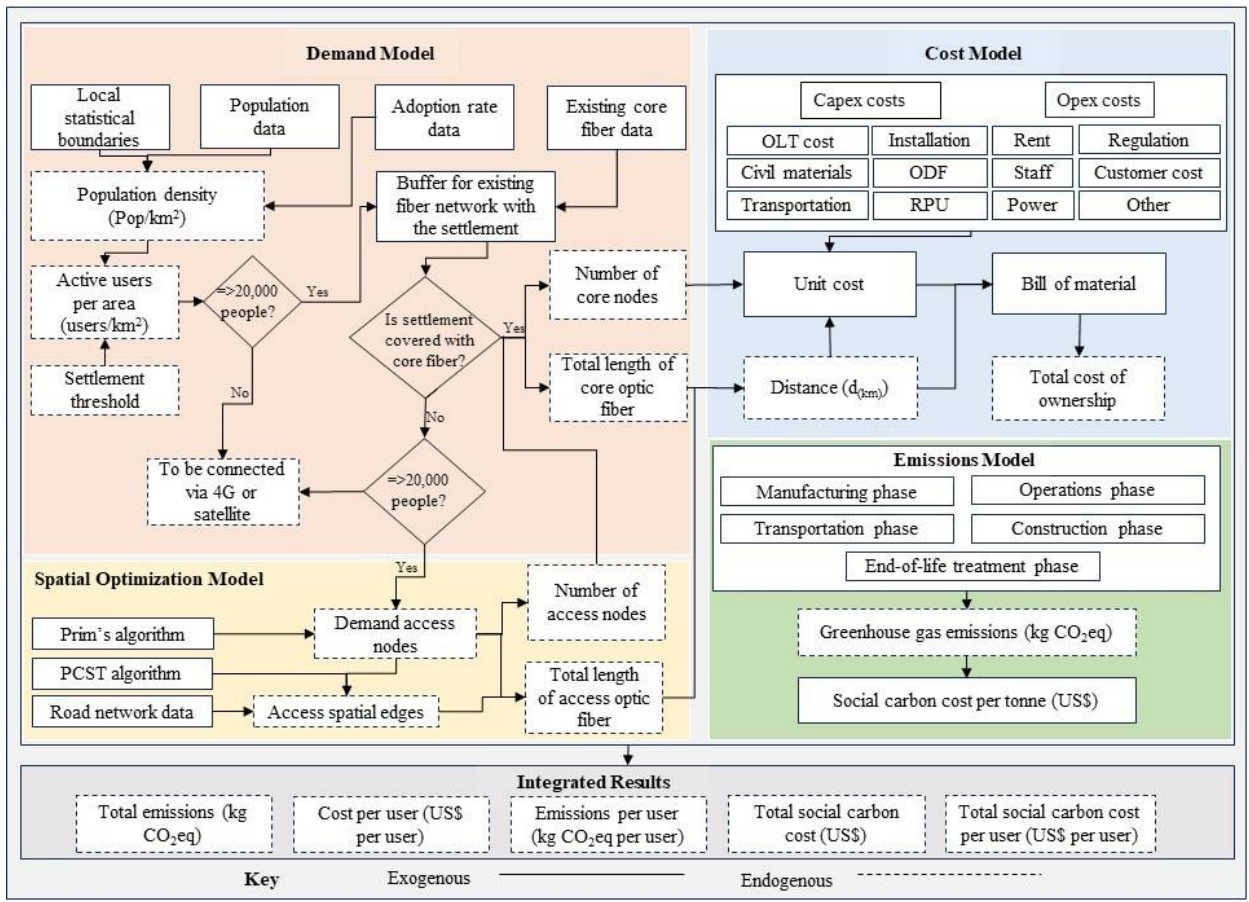

**Fig. 3 | Modeling framework**. Integrated demand, coverage and emissions model for Fiber-to-the-Neighborhood broadband, FTTnb network.

### e. Data and Application

Now that the generalizable system model has been specified, the associated parameters require populating. First, geographic regions must be defined via the Global Administrative Area (GADM) database [88] which contains all administrative areas of every country at several levels of sub-division. GADM uses two levels of country subdivisions including the first level (GID_1) and second level (GID_2). The sub-regions form local statistical areas for installing each fiber node,

FTTnb (see the supplementary information). The GID_1 and GID_2 form the regional and access network level respectively in this model. Next population data is required since it is a major driver of infrastructure investment decisions such as deploying broadband fiber. Data are obtained from WorldPop [89] with the 2020 unconstrained global population mosaic layer used for this study. The population raster layer has a spatial resolution of 30 arc approximating to 1 km at the equator. The raster layer consists of grid cells that contain the number of people, hence the distribution. Next, the main settlement threshold of 20,000 people is set as established in previous studies that seek to deploy broadband in rural areas [9]. The values for the cost model are then obtained from previous studies to estimate the TCO over a 30-year assessment period. The full demand, supply and cost modeling parameters are reported in Table 1.

| Parameter | Unit | Value | Symbol | Source |
|---|---|---|---|---|
| Main settlement population | Inhabitants | 20,000 | - | [46] |
| Fiber node buffer | km | 2 | - | [46] |
| Optical Line Terminal (OLT) | US$ | 28,000 | $C_{OLT}$ | [32] |
| Civil construction materials per terminal node | US$ | 120,000 | $C_{Civil}$ | [46] |
| Transportation | US$ per km | 600 | $C_{Trans}$ | [46] |
| Installation | US$ per km | 6,000 | $C_{Inst}$ | [46] |
| Remote Power Unit (RPU) | US$ | 11,000 | $C_{RPU}$ | [46] |
| Optical Distribution Frame (ODF) | US$ | 18,000 | $C_{ODF}$ | [32] |
| Site rental | US$ | 11,000 | $O_{Rent}$ | [46] |
| Staff and maintenance | US$ | 150,000 | $O_{Staff}$ | [46] |
| Power | US$ | 1,000 | $O_{Pwr}$ | [46] |
| Regulatory fees | US$ | 60,000 | $O_{Reg}$ | [46] |
| Subscriber acquisition | US$ | 120,000 | $O_{Acq.}$ | [46] |
| Other costs | US$ | 180,000 | $O_{other}$ | [46] |
| Discount rate | % | 8.33 | r | [90] |
| Assessment period | Years | 30 | n | [91] |
| Global social carbon cost at 2.5% discount rate | $US per tonne | 75 | $C_{US\$}$ | [87] |

**Table 1 | Modeling parameters.** Empirical demand, supply and cost values applied into the theoretical model.

There are limited data on the existing core fiber networks across SSA. Therefore, we use all the available fiber network data compiled by an open-source provider known as AfterFibre [92]. The data are provided as geographical vector shapefiles for each country. A two-kilometer buffer is created around the existing core fiber network to connect the main settlements, before the largest regional settlement with a population above the set threshold is considered as the main routing node for regional fiber access points.

It is anticipated that the fiber network will be deployed alongside road networks across SSA. The road networks are thus used as the paths for routing the fiber to the settlement nodes. The street data containing all transportation infrastructure are downloaded from Overture Maps Foundation (Overture) [93]. Overture provides open-source geospatial data on buildings, administrative boundaries, places, and transportation networks. The February 2024 transportation data was downloaded to provide the road information across SSA since it is the dominant form of transport.

The LCA data is required for manufacturing, transportation, operation and EOLT as defined by the system boundary diagram in Fig. 2. In the manufacturing phase, the main materials and equipment considered for this analysis are optic fiber cable, other materials, and the housing facility. The optic fiber cable data is based on data from commercial vendors where an outdoor corrugated steel tape and stranded loose tube weighs 247 kg per km [94]. The inventory of other materials are obtained from a previous study that quantified the amount of material needed to build an access point [95]. Table 2 shows all the materials used in the inventory.

The emission values for transporting optic fiber cable and all the listed equipment values from China to the ports of SSA are adopted for international shipping travel, following the method as outlined in a previous country case study for New Zealand [96]. For non-fiber and other materials, the carbon emission factor included in the manufacturing phase accounts for the transportation of the material from the manufacturing plant to the consumer site, as outlined by the UK's Department of Business, Energy and Industrial Strategy (DBEIS) [97].

The power consumption data in the operation phase is based on approximations from previous studies. For instance, the power consumption values for a typical fiber node stations are obtained from past literature [98]. Each of the materials under consideration are assumed to be recycled through an open-loop system where they are converted into new materials or waste. The carbon emission factors for each of the EOLT materials are also presented in Table 2.

| LCA Phase | Metric | Unit | Carbon emission Factor (kg $CO_2$ eq.) | Source |
|---|---|---|---|---|
| **Manufacturing** | Optic Fiber Cable (kg per km) | 247 | 1.403 | [99],[99] |
| | Printed Circuit Board in kg | 3 | 18.76 | [95] |
| | Plastics in kg | 20 | 3.413 | [99],[99] |
| | Steel in kg | 15 | 19.4 | [99], [99] |
| **Transportation** | Optic Fiber Cable | - | 0.3234 | [96] |
| | Other Materials (concrete) | - | 0.3234 | |
| **Construction** | Trenching Distance Percent | 1% | | [100] |
| | Trenching Hours per Distance | 1hr/km | | [100] |
| | Machinery Fuel Efficiency | 24.33 liters/hr | | [100] |
| | Diesel Carbon Emission Factor | 2.68 kg $CO_2$ eq. | | [101] |
| **Operations** | Fiber Distribution Point | 1 | - | [98] |
| | Terminal Node power in kWh | 0.5 | - | [96] |
| | Electricity | - | 0.1934 | [97] |
| **End of Life Treatment (Open – Loop recycling)** | Steel in kg | 15 | 0.9847 | [95] |
| | Optica Fiber Cable (kg per km) | 247 | 2.3 | [82] |
| | Printed Circuit Board in kg | 3 | 18.6 | [95] |
| | Plastics in kg | 20 | 2.3 | [97] |

**Table 2 | LCA Data.** LCA material emission, carbon factor and power consumption data used.

## IV.   Results

### a.   Demand and Supply Results

Table 3 reports the users per square kilometer and the associated characteristics of the geotypes. Even though a significant population of approximately 213 million people live in densely populated areas (>958 persons per km²), about 600 million (52% of the total population) still occupy sparsely populated areas (below 171 persons per km²).

| Geotype | Area(km²) | Population | Percentage of total population (%) | Minimum population density (persons per km²) |
|---|---|---|---|---|
| Decile 1 (>958 per km²) | 77,288 | 212,797,965 | 18.4 | 958 |
| Decile 2 (456 - 957 per km²) | 182,875 | 112,837,064 | 9.8 | 456 |
| Decile 3 (273 - 455 per km²) | 256,765 | 88,018,046 | 7.6 | 273 |
| Decile 4 (172 - 272 per km²) | 664,012 | 139,890,333 | 12.1 | 172 |
| Decile 5 (107 - 171 per km²) | 1,016,325 | 134,491,357 | 11.7 | 107 |
| Decile 6 (64 - 106 per km²) | 1,709,771 | 140,170,957 | 12.1 | 64 |
| Decile 7 (40 - 63 per km²) | 2,438,417 | 121,535,766 | 10.5 | 40 |

| | | | |
|---|---|---|---|
| Decile 8 (22 - 39 per km$^2$) | 3,524,667 | 103,085,613 | 8.9 | 22 |
| Decile 9 (10 - 21 per km$^2$) | 5,387,051 | 76,466,927 | 6.6 | 10 |
| Decile 10 (<9 per km$^2$) | 9,726,874 | 25,472,108 | 2.2 | 1 |
| **Total** | **24,984,045** | **1,154,766,136** | | |

**Table 3 |** Sub-Saharan Africa population geotype characteristics.

In Fig. 4A the population density frequency is presented. The most populous SSA countries (Nigeria, Ethiopia, DRC, Tanzania, South Africa, Kenya, Uganda, Sudan, Angola, and Ghana) contribute some of the highest population density values. In Fig. 4B-C, the spatial distribution of the settlements at sub-regional levels are illustrated. For example, in Fig. 4B we can see that smaller settlement sizes are more numerous, such as below 20,000 inhabitants. We can also view the spatial heterogeneity of medium-sized settlements in SSA, with between 20,000-50,000 people, Fig. 4C. Finally, the best case for infrastructure investment in larger settlements with over 50,000 people as shown in Fig. 4D. These high-density settlement patterns provide a strong motivation for connecting these nodes via fiber deployment. Importantly, the population density distributions reported here provide the basis for estimating the resulting costs, emissions, and SCC.

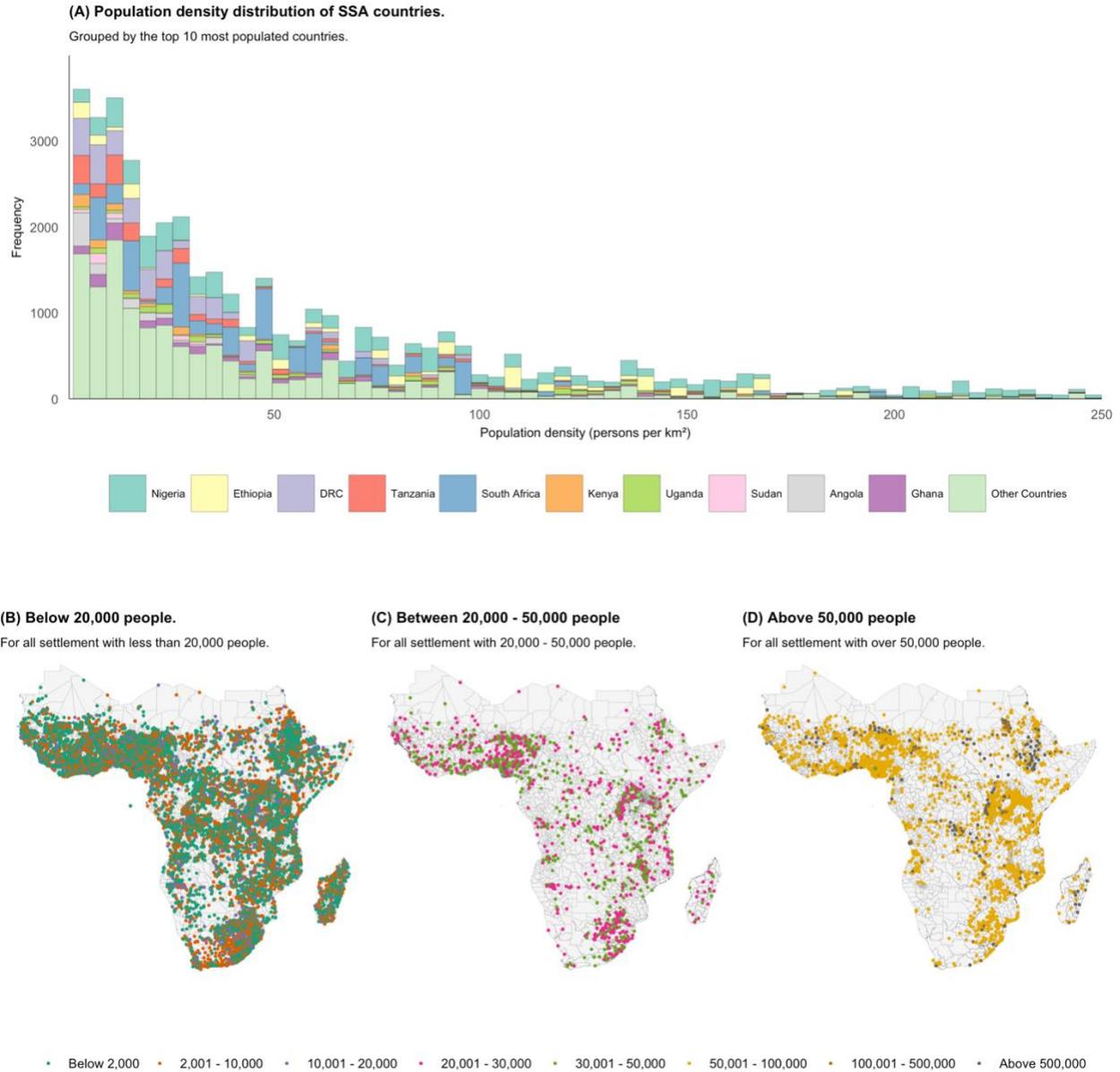

**Fig. 4 | Population results**. **A,** Frequency of the population density considering the top 10 most populated SSA countries. Such information is useful in knowing the countries to prioritize when population is a key driver for fiber deployment project. **B,** Spatial distribution of settlements with population below 20,000 people. **C,** medium sized settlements with 20,000-50,000 people. **D,** high potential settlements with over 50,000 people. A significant portion of Western and Eastern Africa is covered with settlements above 50,000 people.

Next, we report the estimated annualized and monthly TCO per user over the 30-year assessment period. The caveat to these per user TCO results is that they are not the final cost incurred by the fiber broadband users as they will incur additional monthly charges from last mile connectivity service provider. The eventual broadband cost at each premise will vary depending on whether the last mile service is provided by mobile, fixed wireless access, satellite or FTTP.

**(A) Fiber Broadband Total Cost of Ownership (TCO) Reported Per User**
Annualized average TCO per user categorized by deciles, grouped by network level and spatial optimization algorithm.

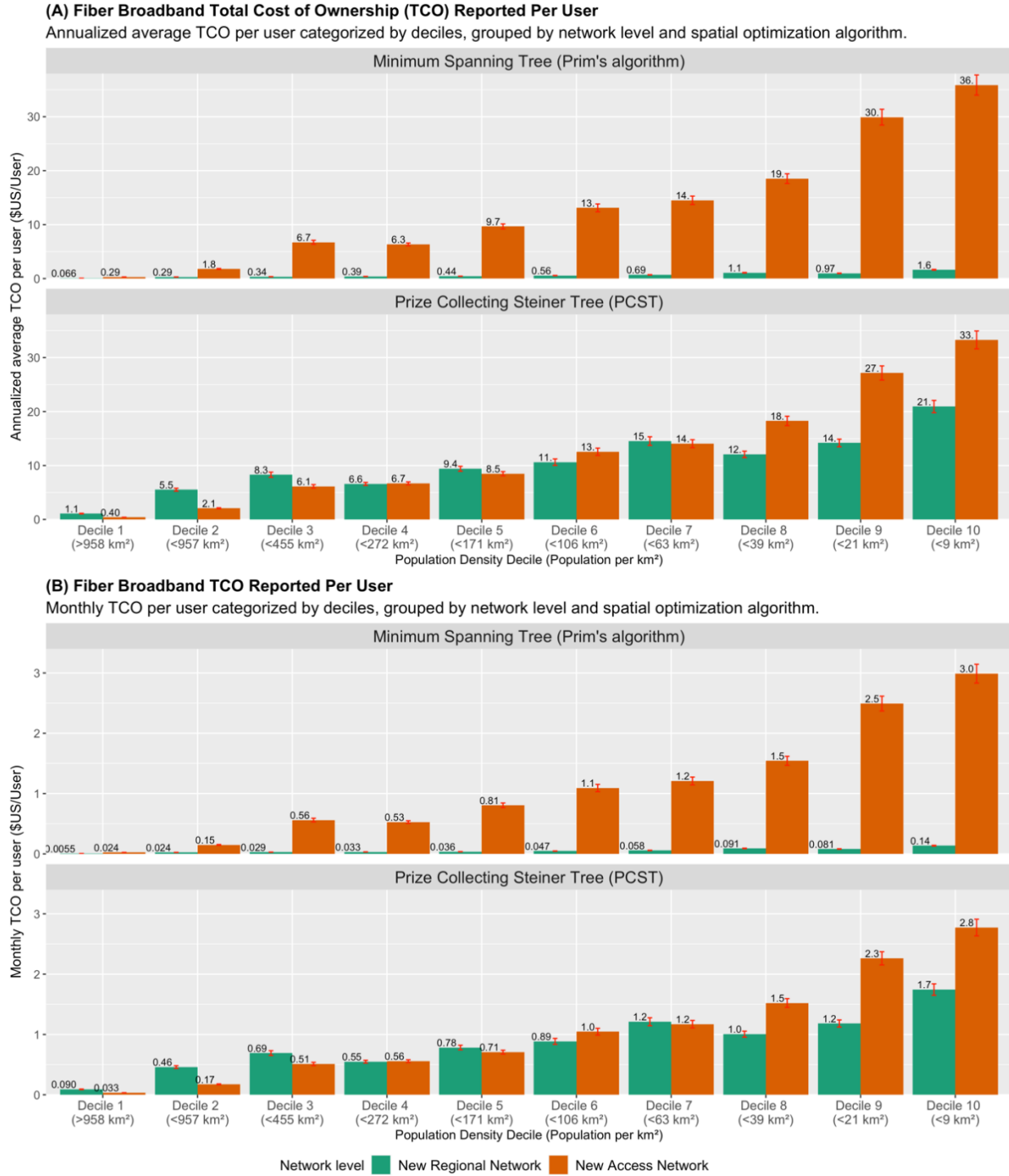

**(B) Fiber Broadband TCO Reported Per User**
Monthly TCO per user categorized by deciles, grouped by network level and spatial optimization algorithm.

**Fig. 5 | Cost results** based on MST Prim's and PCST spatial optimization algorithms at different network levels calculated over the 30-year assessment period. **A,** Average TCO per user reported by geotypes and grouped by network build level. **B,** Monthly TCO per user reported by geotypes and grouped by network build level.

It is important to report the average TCO per user annually since users do not necessarily subscribe to a service for the whole 30-year assessment period, **Fig. 5**A. The annualized TCO per user when the network was designed using MST Prim's algorithm was US$ 1.6 at the regional level and US$ 36 at the access network level in Decile 10. The TCO values were lower for densely populated Decile 1 amounting to US$ 0.066 for regional and US$ 0.29 for the access network. Using a PCST approach resulted in higher costs especially in sparsely populated areas. In Decile 10, the annualized TCO per user was US$ 21 for regional and US$ 33 for the access network. The values were lower for densely populated areas (Decile 1) resulting in US$ 1.1 for regional and US$ 0.4 for the access network.

The mean monthly TCO per user when the network was designed using MST Prim's algorithm amounts to US$ 0.14 (regional) and US$ 2.8 (access) for Decile 10. Likewise, the values for Decile 1 were US$ 0.0055 for regional and US$ 0.024 for the access network, Fig. 5B. A similar trend was observed when using PCST algorithm in designing the network. At regional level the values were US$ 1.7 and US$ 2.8 at access level for Decile 10. As for decile 1, the recorded monthly TCO per user were US$ 0.09 for regional and US$ 0.033 for the access network, Fig. 5B. Even though the per user monthly costs are low, the amount is expected to increase when the charges from the last mile service provider is added.

b. Emission Results

In Fig. 6, we present the carbon emissions results for each of the 10 population geotypes (Decile 1-10) based on the spatial optimization algorithm used in designing the network. In general, deploying fiber in areas with low population density (less than 9 people/km$^2$) such as Decile 10 results in higher average carbon emissions per user. Applying MST Prim's algorithm in constructing the network resulted in 5.5 kg $CO_2$ eq./user (regional) and 55 kg $CO_2$ eq./user (access) of carbon emissions in Decile 10. In contrast, in Decile 1 the average emissions quantity was 0.044 kg $CO_2$ eq./user (regional) and 0.59 kg $CO_2$ eq./user (access). The emissions quantity increased to 2.7 kg $CO_2$ eq./user at the regional level and 3.5 kg $CO_2$ eq./user at the access level in Decile 1 when the PCST algorithm was used to design the network. Estimates were significantly higher for Decile 10 amounting to 129 kg $CO_2$ eq./user (regional) and 287 kg $CO_2$ eq./user (access).

The annual emissions per user are also presented in Fig. 6B. Using MST Prim's algorithm, the recorded emissions in Decile 10 are 0.18 kg $CO_2$ eq./user for regional and 1.8 kg $CO_2$ eq./user for the access network. For densely populated areas (over 958 people/km$^2$), the values decrease on a per user basis. Indeed, the reported values were 0.0015 kg $CO_2$ eq./user (regional) and 0.02 kg $CO_2$ eq./user (access) in Decile 1, Fig. 6B. Applying PCST spatial optimization results in higher annualized per user emissions. In Decile 10, the emissions were 4.3 kg $CO_2$ eq./user for regional and 9.6 kg $CO_2$ eq./user for access, as per Fig. 6B. For densely populated areas such as Decile 1, low annualized emissions of 0.088 kg $CO_2$ eq./user are estimated at the regional level and 0.12 kg $CO_2$ eq./user at the access network level. To conclude, elevated emissions values are found in sparsely populated areas (less than 9 people/km$^2$).

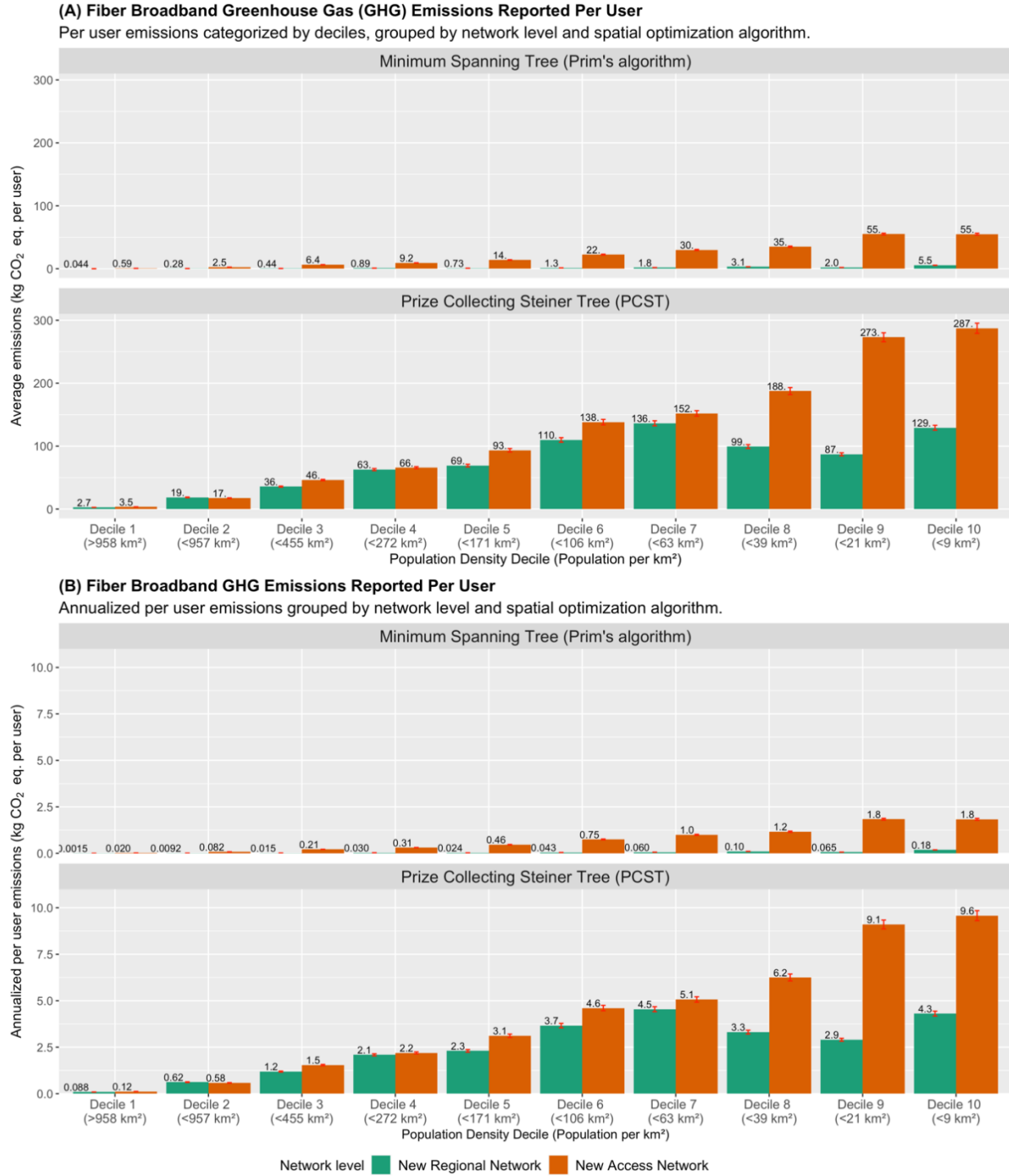

**Fig. 6 | Emission results. A,** Average per user GHG emission results based on fiber design using MST Prim's and PCST algorithm reported by geotypes and categorized by network build level. **B,** Annualized per user GHG emission results based on fiber design using MST Prim's and PCST algorithm reported by geotypes and categorized by network build level.

The monetary value of environmental damages caused to society are estimated and presented in terms of SCC (Fig. 7). As was the case with average emissions per user, the resulting SCC per user was higher when the PCST algorithm was applied to design the network compared to the MST Prim's algorithm. Moreover, sparsely populated areas (less than 9 people/km$^2$) recorded the highest SCC per user. For instance, in Decile 10, the average SCC considering the regional network was US$ 0.41 and US$ 4.1 per user for the access network over the 30-year period, Fig. 7A. In comparison, the fiber network design in Decile 1 resulted in US$ 0.0033 (regional) and US$ 0.044 (access) SCC per user (Fig. 7A). These SCC estimates are higher when the PCST algorithm was used to design the network. The SCC per user in sparsely populated areas (less than 9 people/km$^2$) such as Decile 10 (US$ 9.7 for regional and US$ 22 for access network) is significantly higher. Densely populated areas (over 958 people/km$^2$), such as Decile 1, see the SCC per user drop to US$ 0.2 for regional infrastructure and US$ 0.26 for the access network.

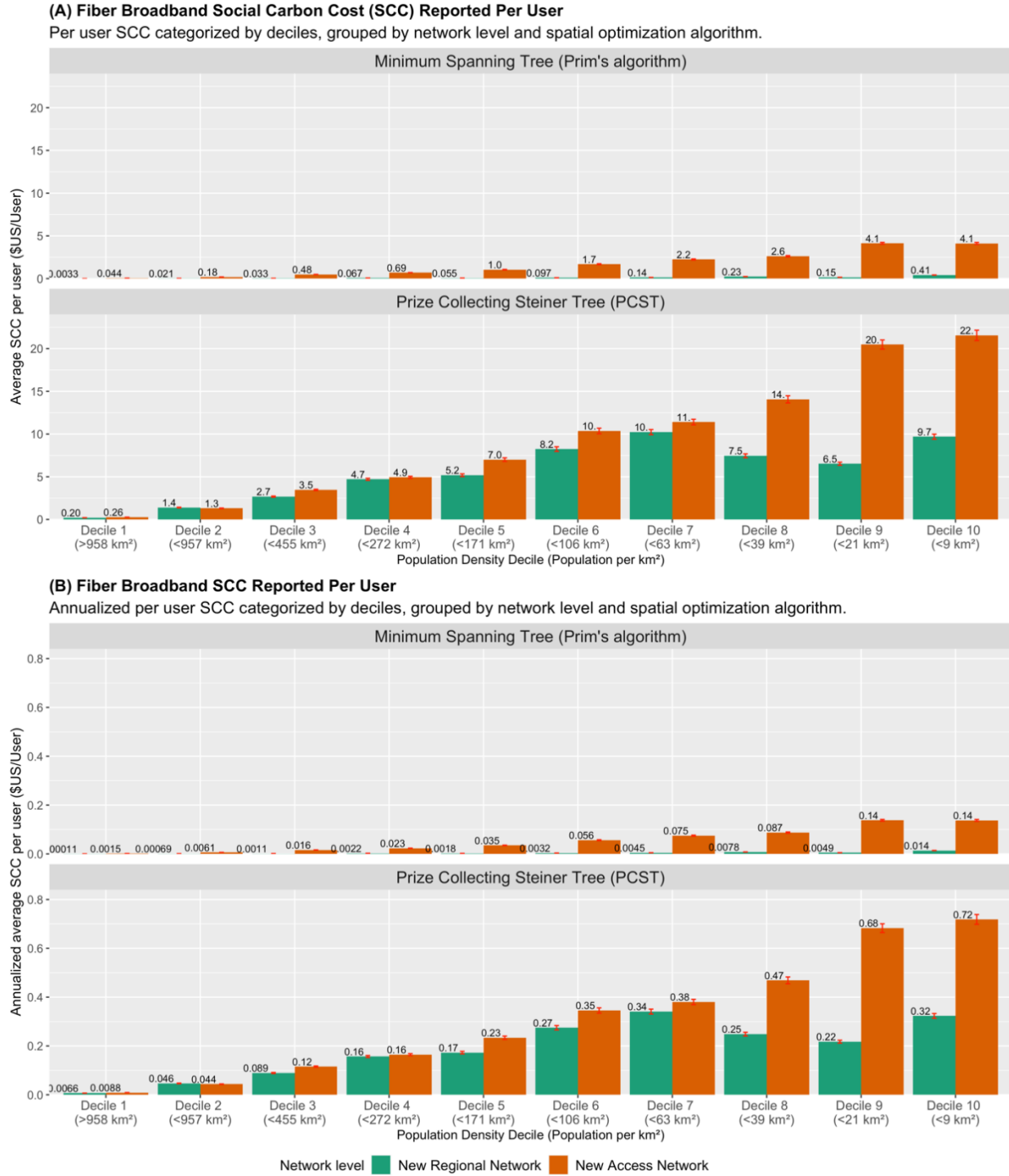

**Fig. 7 | Social Carbon Cost (SCC)** results based on MST Prim's and PCST spatial optimization algorithms at different network levels over the 30-year assessment period. **A,** SCC per user calculated based on MST Prim's and PCST algorithm. **B,** Annualized SCC per user calculated based on MST Prim's and PCST algorithm.

Next, the SCC per user is broken down annually and presented in Fig. 7B. The SCC is lower when using MST Prim's algorithm compared to PCST. Also, the SCC is lower in densely populated areas (over 958 people/km$^2$) such as Decile 1 compared to sparsely populated areas (Decile 10). The SCC values for the regional and access network are approximately US$ 0.0011 (regional) and US$ 0.0015 (access) when the fiber line in Decile 1 is designed using MST Prim's algorithm. The value increased to US$ 0.014 for regional and US$ 0.14 for the access network in sparsely populated Decile 10. However, designing the network using PCST results in higher SCC per user of US$ 0.32 (regional) and US$ 0.72 (access) in Decile 10. The amounts are comparatively lower for densely populated areas such as Decile 1 (US$ 0.0066 for regional and US$ 0.0088 for access network).

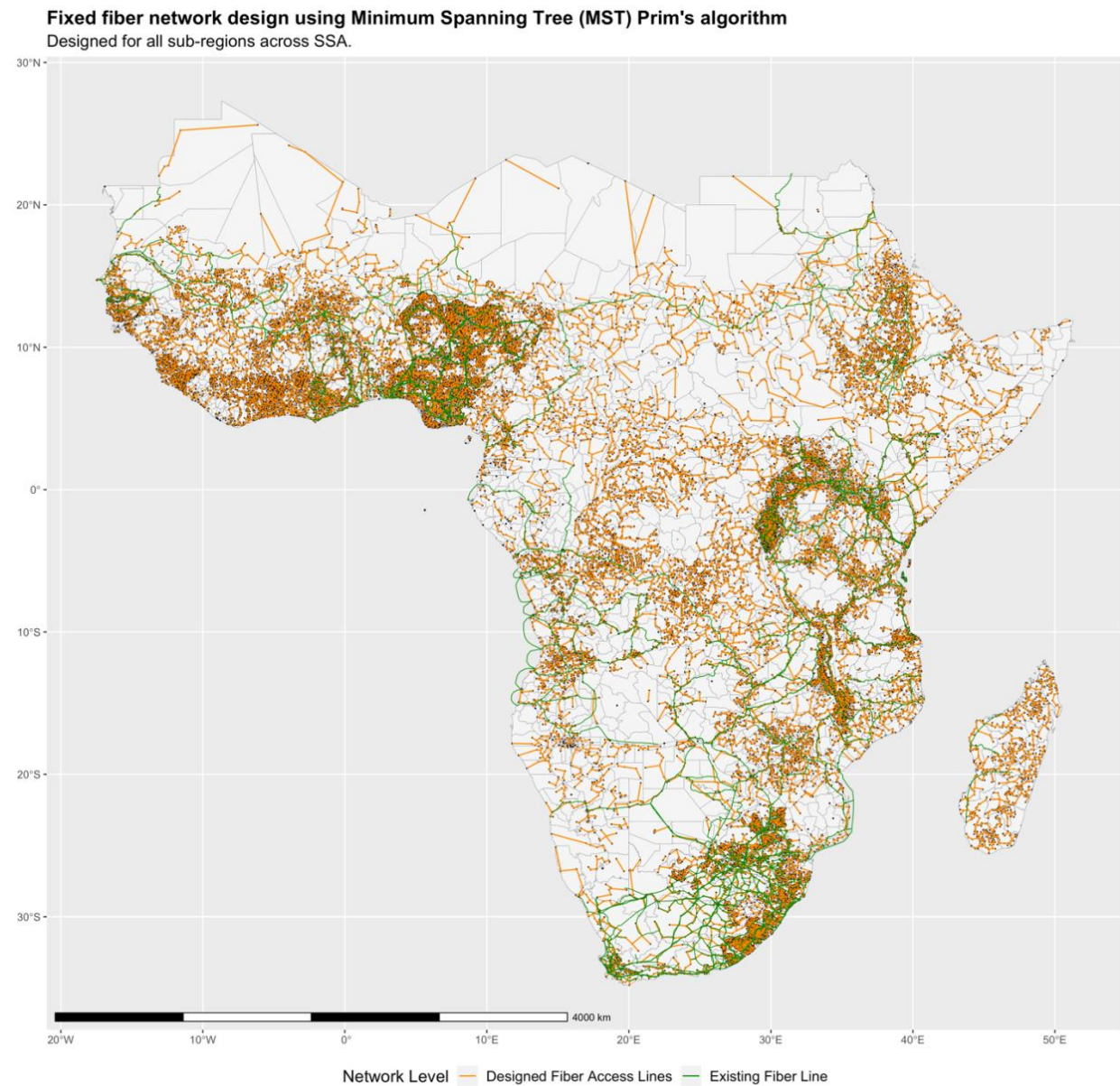

**Fig. 8 | Fiber design.** Fixed fiber network design using MST Prim's algorithm.

In **Fig. 8**, the fiber network designed using MST Prim's algorithm is visualized. Via the MST approach, the number and total distance of the fiber lines is lower compared to PCST. Also, fewer terminal nodes are required. However, for the MST algorithm each of the terminal nodes are connected by a fiber line. This is due to the inherent property of MST algorithms, which mathematically seek to minimize the cost of connecting all the nodes in the networks. As a result, each of the nodes are connected via a least-cost approach, although the downside is that the distance connecting the nodes is not necessarily realistic as the algorithm calculates the Euclidean distance between the nodes.

**Fixed fiber network design using Prize Collecting Steiner Tree (PCST) algorithm**
Designed for all sub-regions across SSA.

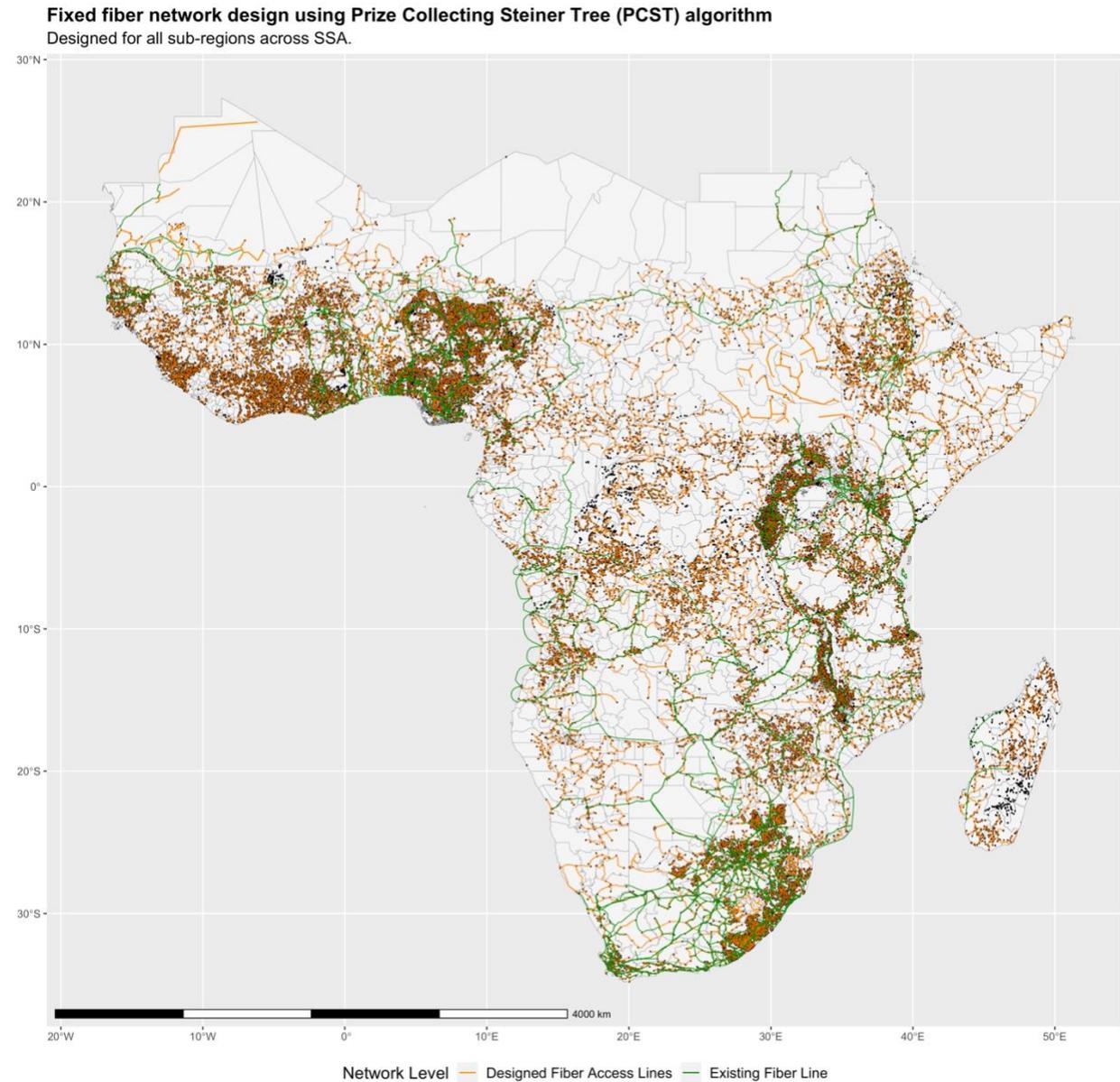

Network Level  ── Designed Fiber Access Lines  ── Existing Fiber Line

**Fig. 9 | Fiber design.** Fixed fiber network design using Prize Collecting Steiner Tree (PCST) algorithm.

To design a realistic least-cost network where the fiber line can actually be routed, the PCST algorithm is applied utilizing road data. In **Fig. 9,** the least-cost fiber network designed using the PCST algorithm is shown. Unlike MST, the PCST covers a greater distance thus requiring more fiber to be built. Although in contrast, of the total input terminal nodes supplied to the PCST algorithm, some are not included in the final tree network. For instance, the average number of access terminal nodes for PCST was 119 compared to 158 from MST design. Again, this is due to the nature of the algorithm. Mathematically, PCST seeks to find the least-cost path between vertices in a graph network, achieving this objective by sometimes ignoring certain vertices (and incurring a penalty for each unconnected vertex in the final network). This is well demonstrated in **Fig. 9,** and visible in the Northern countries of SSA such as Chad, Mauritania, Niger and Sudan where some nodes remain unconnected.

## V.    Discussion

Having presented the results, a discussion is now undertaken with regard to the implications for the research questions previously articulated

*What is the best way to model fixed broadband infrastructure networks using spatial optimization algorithms to reduce investment costs and emissions?*

In this study, we have quantified the cost and emissions associated with deploying fixed fiber broadband across 44 countries in SSA using two spatial optimization algorithms. The aggregate investment cost of deploying FTTnb infrastructure in sparsely populated areas (<9 people/km$^2$) is US$ 25-26 billion (16-22 times higher than densely populated areas) (see supplementary information). Similarly, the resulting carbon emissions of deploying FTTnb in sparsely populated areas (<9 people/km$^2$) is 17 times higher (when using MST to design access network) and 14 times higher (when using PCST to design the access infrastructure) than densely populated areas (over 958 people/km$^2$). These values equate to 0.081 Mt $CO_2$ eq. in Decile 1 and 1.4 Mt $CO_2$ eq. in Decile 10 (MST access network). For the PCST access network, the values are 0.49 Mt $CO_2$ eq. (Decile 1) and 7.1 Mt $CO_2$ eq. (Decile 10). Our study provides crucial insight on the relationship between costs and carbon emissions for the deployment of fiber infrastructure, with results disaggregated at granular population density levels.

The analysis is broadly in concordance with previous assessments [53], [80], [102] with regard to the estimated costs and emissions on a per user basis, with lower population density areas (<9 people/km$^2$) incurring much higher emissions. For example, deploying FTTnb by applying PCST algorithm to areas with less than 9 people/km$^2$ increases annual emissions by approximately 49 times (regional network) and 80 times (access network) equivalent to 4.3 kg $CO_2$ eq./user (regional) and 9.6 kg $CO_2$ eq./user (access) when compared to urban areas exceeding 958 people/km$^2$. The values are comparatively lower when the MST algorithm is used translating to 0.18 kg $CO_2$ eq./user (regional network) and 1.8 kg $CO_2$ eq./user (access network) thus 12 and 90 times higher than densely populated areas respectively.

Ideally, if cost was not an important factor, then network operators would try to maximize the number of buildings served by FTTP, as this architecture provides superior capacity, low latency and robust reliability. However, even when utilizing a less ideal architecture, such as FTTnb, the costs and emissions results for the SSA region quickly break down in sparsely populated areas (e.g., Deciles 6-10). Indeed, a spatial analysis of SSA population density shows that 52% of the total population are living in areas with a population density of below 106 people/km$^2$. To this end our results demonstrate that for urban areas with high population density (>958 people/km$^2$), the annualized TCO per user is 96-99% (US$ 0.066-0.29) lower compared to sparsely populated

regions (<9 people/km$^2$) when using an MST algorithm. This compares to the PCST method where the annualized TCO is 95-99% (US$ 0.4-1.1) lower. The caveat to these cost values is that they are not inclusive of the last mile connectivity charges to the end user. For the large geographic areas assessed in this analysis, the results suggest that building this quantity of fiber network infrastructure is expensive irrespective of the spatial optimization algorithm used. Over the next decade, these areas will continue to be served by cheaper deployment methods, utilizing more long-distance wireless radio links to reduce cost.

*How much investment is required in fixed broadband infrastructure to achieve affordable universal connectivity, for example, focusing on a Fiber-to-the-Neighborhood (FTTnb) approach?*

We have estimated that the investment cost of deploying FTTnb in remote areas (<9 people/km$^2$) of SSA using MST algorithm is about US$ 26 billion. The cost is comparatively lower (US$ 25 billion) when the PCST algorithm is used (see supplementary information). To put these figures into context, the 2023 Gross Domestic Product (GDP) value for SSA was US$ 2.03 trillion [103]. In other words, building a fixed fiber broadband infrastructure close to settlement points across SSA will cost 1.28% of annual GDP when using MST and 1.23% when applying PCST.

It is important to compare these values at per user levels to avoid drawing incorrect conclusions. The average TCO per user in sparsely populated areas (<9 people per km$^2$) is 17-25 times higher (US$ 33-36) compared to densely populated regions (>958 people/km$^2$) depending on the spatial optimization algorithm used in the FTTnb network design. Operators are unlikely to build fiber broadband infrastructure in such areas unless other interventions such as government support are adopted. One option available to network operators is the use of cross-subsidies, for example by raising the installation and monthly subscription charges in densely populated areas, to offset the high cost per user in remote areas. This is frequently how mobile cellular networks are operated, in so far as only a fraction (e.g., 20%) of the sites might be profitable overall. Yet, the raison d'être of the network is to provide mobility, forcing operators to cross-subsidize infrastructure deployment in rural areas to ensure coverage. Granted, this may be unpalatable for operators in price competitive markets.

In the case of SSA, FTTnb approach makes a logical business case to bring fiber closer to the end users in a portion of areas (48% of the total SSA population). However, our results shows that even in the case of FTTnb, the approach may be unviable in some places with lower than 106 people/km$^2$ (Decile 6-10). In these areas, the annualized TCO per user is US$ 13 (MST access) and US$ 33 (PCST access), translating to 64% for MST and 65% for PCST higher, without the cost of the last-mile connectivity option. In densely populated areas with over 106 people/km$^2$ (Decile 1-5), the TCO per user ranges between US$ 0.29 (MST access) to US$ 8.5 (PCST access) that is equivalent to 85% and 97% higher respectively. This variance justifies the rational need for operators to build FTTnb to the most populated areas first before focusing on sparsely populated regions (less than 9 people/km$^2$). The remaining remote areas can be connected using other alternatives, such as satellites, at designated community access points.

*What are the environmental sustainability implications of broadband infrastructure strategies?*

Our modeling approach considered two spatial optimization algorithms to reduce the cost and carbon emissions in building fiber closer to users across 44 SSA countries. We have established that designing a fiber network using the MST algorithm in sparsely populated areas (less than 9 people/km$^2$) results in 0.14 Mt $CO_2$ eq. (Decile 10 regional) and 1.4 Mt $CO_2$ eq. (Decile 10 access). The Decile 10 total emission values are 2 (regional) and 17 (access) times higher compared to Decile 1 densely populated areas (over 958 people/km$^2$). Similarly, the emissions are 9 (regional) and 14 (access) times higher in Decile 10 compared to Decile 1. The relative values equate to 3.2 Mt $CO_2$ eq. (Decile 10 regional) and 7.1 Mt $CO_2$ eq. (Decile 10 access) higher than densely populated areas (over 958 people/km$^2$) when the network is designed using the PCST algorithm. To illustrate the implications of these results, the World Bank reported 834 Mt $CO_2$ eq. of emissions across SSA in 2019 prior to the COVID-19 pandemic [104]. That is to say that building a fixed fiber broadband infrastructure close to settlement points across SSA results in 0.01-1.7% of cross sector carbon emissions in SSA when using an MST approach and 0.04-0.85% when applying a PCST approach.

Importantly, the interplay of the provided capacity, required investment costs and resulting emissions make for interesting reading, as we are not aware of other studies which have carried out integrated assessment for SSA. We find that the quantity of emissions per user in sparsely populated areas (Decile 10) is 5.5 kg $CO_2$ eq./user (regional) and 55 kg $CO_2$ eq./user (access) when using MST algorithm (93-125 times higher than densely populated areas). When using the PCST algorithm, the per user emissions in Decile 10 is 129 kg $CO_2$ eq./user (regional) and 287 kg $CO_2$ eq./user (access) which is 48 and 82 times higher than densely populated areas (Decile 1) respectively. As with costs, low population densities mean the emissions incurred from building infrastructure assets need to be split over fewer users, increasing per user emissions. For instance, in areas with less than 9 people/km$^2$(Decile 10), the resulting annualized per user emissions when using the MST algorithm is 0.18 kg $CO_2$ eq./user (regional) and 1.8 kg $CO_2$ eq./user (access) (12 and 90 times higher compared to regions with over 958 people/km$^2$). Similarly, the Decile 10 annualized per user emissions is 4.3 kg $CO_2$ eq./user (regional) and 9.6 kg $CO_2$ eq./user (access) (49 and 80 times higher) when the PCST algorithm is used to design the fixed fiber network.

Finally, it is important to measure the monetary damage caused by carbon emissions resulting from building fixed fiber broadband using the SCC metric. In this study we estimated that the total SCC in sparsely populated areas (<9 people/km$^2$) across the 44 SSA countries is US$ 0.01-0.53 million (2-16 times higher than densely populated areas). Such high SCC values in the rural and remote areas underscore the environmental cost of deploying fixed fiber infrastructure. Currently, fiber is fronted as potentially the greenest form of broadband connection over long time horizons, compared to other methods such as fixed wireless or cellular broadband. However, using the SCC as a metric for quantifying the monetary cost damage of emissions, deployment of fixed fiber broadband in less populated areas results in elevated carbon emissions equating to US$ 0.41-22 per user (12-93 times higher compared to areas with over 958 people/km$^2$).

## VI.     Conclusion, limitations, and further research

In this study we quantify the investment costs, emissions, and SCC for fiber broadband at the sub-national level across 44 SSA countries. The use of global population and geospatial boundary data enables the quantification of these three metrics to evaluate FTTnb viability. The cost of using the MST algorithm to design the fixed fiber network is approximately US$ 26 billion (1.28% of the SSA annual GDP). On the other hand, the total investment cost when the PCST algorithm is used is US$ 25 billion, accounting for 1.23% of SSA's GDP. The key take-away is that despite taking a more modest approach than FTTP, providing fiber broadband through FTTnb is not viable in many areas (52% of the total population), and likely only possible over the next decade in the first five population deciles (Decile 1-5), covering ~550 million people. For perspective, only 48% of the total SSA can likely be viably connected using this the FTTnb approach in the next ten years. Thus, network operators and governments need to work together to creatively lower the costs of deployment, as well as encouraging business model innovation to push out coverage, such as via infrastructure sharing.

We quantify the degree to which FTTnb is associated with higher emissions in lower population density areas. For example, on average the annualized per user emissions is about 0.18-1.8 kg $CO_2$ eq./user (when an MST algorithm is used) and 4.3-9.6 kg $CO_2$ eq./user (when a PCST algorithm is used). These values are higher compared to the annual emissions per user in densely populated areas (over 958 people/km$^2$). For instance, when using an MST algorithm, the emissions are 0.015-0.02 kg $CO_2$ eq./user (12-90 times lower) and 0.088-0.12 kg $CO_2$ eq./user (49-80 times lower) when using PCST algorithm. The emission difference leads to per user monetary damage costs of US$ 0.41-4.1 in sparsely populated areas (<9 people/km$^2$) when using an MST algorithm and US$ 9.7-22 for PCST algorithm. In densely populated areas (>958 people/km$^2$), the per user SCC is US$ 0.033-0.044 (MST) and US$ 0.2-0.26 (PCST). That is to say that the environmental and social damage cost of deploying fixed fiber broadband in remote/rural areas (less than 9 people/km$^2$) is 12-93 times (MST) and 49-85 times (PCST) compared to urban areas (over 958 people/km$^2$).

Lastly, there are limitations to the model developed in this study, such as the uncertainty associated with the quantity and cost of materials used in the LCA and cost model. Although a Monte Carlo simulation approach has been applied to generate a range of estimated values, uncertainty lowers the accuracy of the end results. Additionally, the LCA system boundary here did not include emissions from the extraction and acquisition of raw materials, instead considering this quantity implicitly within the existing emission factors, affecting the total emissions quantities and resulting SCC estimates.

In the future it would be beneficial for this study to be revisited with more accurate data on the cost and quantity of materials from network operators. Future research should also consider applying the same approach but including the final last-mile access network, to compare the results with those reported in this study. A cross-technology comparison provides policy makers with a range of options on how best to connect the unconnected population, while lowering investment costs and associated emissions.

## Data and Code Availability

The full geospatial broadband model developed in Python and R programming language used in generating the results and the accompanying datasets utilized in the paper are available through this GitHub link.

## Funding Statement

This research was funded by Pozibl Inc (Ex Hexvarium).

## Conflict of Interest

The authors have no conflicts of interest to report.